\documentclass[10pt, a4paper]{article}
\usepackage{lrec}
\usepackage{graphicx}
\usepackage{tabularx}
\usepackage{soul}

\usepackage{epstopdf}
\usepackage[latin1]{inputenc}

\usepackage{hyperref}
\usepackage{xstring}
\usepackage{caption}

\usepackage{xcolor}
\usepackage{danudefs}
\usepackage{booktabs}
\usepackage{xifthen}

\usepackage{todonotes}

\makeatletter
\if@todonotes@disabled

\else

\fi
\makeatother

\makeatletter
\newcommand{\ssymbol}[1]{^{\@fnsymbol{#1}}}
\makeatother

\newcommand{\SM}{$\textrm{CoQAM}_{\textrm{SM}}$ }
\newcommand{\MP}{$\textrm{CoQAM}_{\textrm{MP}}$ }
\renewcommand{\SS}[1][]{ \ifthenelse{\equal{#1}{}}{$\textrm{CoQAM}_{\textrm{SS}}~$}{$\textrm{CoQAM}_{\textrm{SS}}(\textrm{#1})$} }

\title{\textit{Do not let the history haunt you} -- Mitigating Compounding Errors in Conversational Question Answering}

\name{Angrosh Mandya$^*$ James O'Neill$^*$, Danushka Bollegala$^{*,\dagger}$, Frans Coenen$^*$}

\address{University of Liverpool$^*$, Amazon$^\dagger$ \\
         \{angrosh,James.O-Neill,danushka,coenen\}@liverpool.ac.uk\\}

\abstract{
The Conversational Question Answering (CoQA) task involves answering a sequence of inter-related conversational questions about a contextual paragraph. 
Although existing approaches employ human-written ground-truth answers for answering conversational questions at test time, in a realistic scenario, the CoQA model will not have any access to ground-truth answers for the previous questions, compelling the model to rely upon its own previously predicted answers for answering the subsequent questions.
 In this paper, we find that compounding errors occur when using previously predicted answers at test time, significantly lowering the performance of CoQA systems. 
 To solve this problem, we propose a sampling strategy that dynamically selects between target answers and model predictions during training, thereby closely simulating the situation at test time.
 Further, we analyse the severity of this phenomena as a function of the question type, conversation length and domain type. 
\\ \newline \Keywords{Scheduled Sampling, Conversational Question Answering, CoQA}}

\begin{document}

\maketitleabstract

\section{Introduction}
\label{sec:intro}

Recently, there has been a significant interest in developing systems that are able to understand text passages and answer a series of inter-connected questions that appear in a conversation~\cite{reddy2018coqa,choi2018quac,yatskar2018qualitative,huang2018flowqa,zhu2018sdnet}. 
To this end, the Conversational Question Answering (CoQA) challenge\footnote{\label{footnote_coqa}https://stanfordnlp.github.io/coqa/} was designed to facilitate development of CoQA systems by providing access to the CoQA dataset, a large-scale dataset comprising of conversational questions and answers \cite{reddy2018coqa}. 
The rapidly increasing list of systems on the CoQA Leaderboard compete with each other to achieve higher performance, clearly demonstrates that machines are quite capable in answering conversational questions. 
In fact, the \textbf{RoBERTa+AT+KD (ensemble)}~\cite{ju2019technical}, the top-ranked\footnote{As at November, 2019.} CoQA model listed on the CoQA Leaderboard, outperforms human-level performance, indicating the superiority of CoQA systems.

An important aspect of CoQA systems is the usage of conversational history as a salient feature for answering conversational questions \cite{reddy2018coqa,yatskar2018qualitative,huang2018flowqa}. 
For example, in the conversational passage shown in Figure~\ref{figure:sample_coqa}, in order to answer $Q_3$ correctly, it is necessary to know the answers from previous questions $Q_1$ (``white'') and $Q_2$ (``in a barn''), without which the system would easily fail due to the absence of useful contextual information. A similar pattern follows in consecutive questions, where information from past questions becomes absolutely vital in answering the current question.

\begin{table}[t]
\begin{center}
{\small
\begin{tabular}{p{7.2cm}}
\toprule
Once upon a time, in a barn near a farm house, there lived a little white kitten named Cotton. Cotton lived high up in a nice warm place above the barn where all of the farmer's horses slept. But Cotton wasn't alone in her little home above the barn, oh no. She shared her ... \\
\\
$Q_1:$ What color was Cotton? \\
$A_1:$ white\\
\\
$Q_2:$ Where did she live? \\
$A_2:$ in a barn \\
\\
$Q_3:$ Did she live alone? \\
$A_3:$ No \\
\\
$Q_4:$ Who did she live with?\\
$A_4:$ with her mommy and 5 sisters \\
\\
$Q_5:$ What color were her sisters?\\
$A_5:$ orange and white \\
\bottomrule
\end{tabular}
}
\end{center}
\captionof{figure}{\label{figure:sample_coqa}Example conversation from CoQA Dataset}
\end{table}


Although, competent systems are developed for the CoQA task, almost all of the current systems \cite{reddy2018coqa,huang2018flowqa,zhu2018sdnet,yatskar2018qualitative} employ conversational history for predicting answers. 
The main drawback of these systems is that they use ground-truth answers, written by human annotators, both during training and test time. 
However, in a real-world QA application, although the model has access to previous questions, access to ground-truth answers in the conversational history is not available to the model at test time. 
However, past history cannot be ignored in CoQA and a more realistic approach is to use model's own predicted answers for the previous questions in the current conversation to answer as the contextual history for the current question. 
However, if the model is trained using the standard maximum likelihood estimate (MLE) with full supervision, it can result in a severe \emph{exposure bias}~\cite{bengio2015scheduled}, making the model unable to recover from its own past errors during test time. 
Moreover, the incorrectly predicted answers for the current question is likely to induce further errors when answering future questions because the former is used in the process of answering the latter.
Therefore, the CoQA model will compound errors along the sequence of answering questions, similar to the well-known problem in sequence prediction tasks when using \textit{teacher forcing}~\cite{williams1989learning}, where only the target labels are used during training.

Despite the need to carefully consider and mitigate exposure bias in CoQA systems, to the best of our knowledge, this problem has not been studied in prior work in the context of CoQA.
On the contrary, the focus so far in current CoQA systems~\cite{reddy2018coqa,huang2018flowqa,yatskar2018qualitative,zhu2018sdnet} has been centred around reporting superior performance on the designated leaderboard.
The official evaluation protocol allows the systems to use ground-truth answers to past questions during test time, thereby conveniently sweeping the exposure bias under the carpet.
Indeed the experimental results (shown later in \autoref{table:overall_results}) presented in this paper clearly show a significant decrease in a CoQA system's overall performance when the model's predicted answers for previous questions are used in place of ground-truth answers for previous questions.

In this paper we propose a variant of \emph{Scheduled Sampling} (SS)~\cite{bengio2015scheduled} to address the compounding errors issue in CoQA systems. 
Specifically, SS uses a mixture of both (a) ground-truths, and (b) model's own predictions during training according to a gradual sampling schedule that shifts from (a) to (b).
SS was originally proposed to overcome the exposure bias in text generation tasks such as machine translation~\cite{bengio2015scheduled} and image captioning~\cite{zhang2017actor}. 

In the context of CoQA, we propose a sampling schedule that uses both ground-truth answers and a model's own predictions for training the CoQA model, which is then used to predict answers without using ground-truth answers during inference. 
Instead of sampling between current predictions and targets, our variant chooses between the previous epoch predictions and current target, whereby the previous predictions are stored. Unlike scheduled sampling, this method can be used with current CUDA optimized recurrent neural networks for popular modules in deep learning frameworks (e.g nn.RNN/GRU/LSTM modules in PyTorch).


We emphasise that the main purpose of this research is \emph{not} to develop a CoQA system that outperforms the state-of-the-art on CoQA, but to investigate the exposure bias evident in CoQA task. 
Instead our objective is to evaluate whether SS can be used to mitigate the compounding errors in CoQA.
As a specific instance of a CoQA system, we select SDNet~\cite{zhu2018sdnet} and evaluate the effect of exposure bias on it.
Specifically, in this paper we address the following research questions:
\begin{enumerate}
    \item \emph{What is the discrepancy in development set performance when we replace previous ground truth answers with the predicted ground truth answers?}
    \item \emph{If there is a significant discrepancy, can well-known methods for mitigating exposure bias, such as the aforementioned SS reduce this performance gap?}
\end{enumerate}
The main contributions of this paper are: 
(a) empirical results that highlight the effects of exposure bias in CoQA and a thorough analysis thereof (i.e., evaluate performance when ground-truth answers are replaced with predicted answers in CoQA models) and,
(b) propose and evaluate various SS techniques to overcome exposure bias in CoQA systems.

\section{Related Work}
\label{sec:related}

The \textbf{Co}nversational \textbf{Q}uestion \textbf{A}nswering (CoQA) dataset \cite{reddy2018coqa} was developed for evaluating systems for answering questions in the form of a continuous dialogue. 
The key characteristic of this dataset is that it provides human-like conversational questions and preserves the naturalness of answers present in conversations. Several baseline systems such as sequence-to-sequence (Seq2Seq)~\cite{mRNN}, pointer-generator network (PGNet)~\cite{See_2017}, Document Reader Question Answering (DrQA)~\cite{chen-EtAl:2017:Long4} and a combined DrQA+PGNet model were initially proposed for CoQA \cite{reddy2018coqa}. 
The BiDAF++ model \cite{yatskar2018qualitative}, based on the Bi-directional attention flow (BiDAF)~\cite{seo2016bidirectional}, used self-attention \cite{clark2017simple} to answer conversational questions. 
\textsc{FlowQA}~\cite{huang2018flowqa} used a mechanism to incorporate intermediate representations generated during the process of answering previous questions, through an alternating parallel processing structure. 

In comparison to previous approaches that simply concatenated previous questions/answers as the input, the intermediate representations provided a more deeper representation of the conversational history. 
The SDNet~\cite{zhu2018sdnet} model used both inter-attention and self-attention to comprehend conversation context and extract relevant information from passage to answer conversational questions. 
More recently, the RoBERTa+AT+KD model~\cite{ju2019technical} is proposed for CoQA, which combines rational tagging multitask, adversarial training, knowledge distillation and a linguistic post-processing strategy to achieve the state-of-the-art performance for CoQA. 
In addition to the above systems, models based on BERT~\cite{devlin2018bert} such as Google SQuAD 2.0 + MMFT, ConvBERT and BERT+MMFT+ADA are also shown to be highly effective for CoQA. 
The XLNet~\cite{yang2019xlnet} model which adopts a generalised autoregressive pretraining method to learn bidirectional contexts and overcomes the limitations of BERT resulting from using masked inputs, is shown to achieve higher performance in comparison to systems that are based on BERT.

Interestingly, the systems described above predominantly use previous questions and answers to model a conversational context for answering a conversational type question. 
Furthermore, the above systems also use ground-truth answers available for previous questions to answer a given question at test time. 
However, in a realistic scenario, such ground-truth answers (for previous questions) is not available for answering a particular question. 
Thus, the system has to either ignore previous answers or use the previous answers predicted by the system itself to answer the current question. 
In both cases, it is likely to hurt the systems performance due to the absence of conversational history. 
Given this problem, this study focuses on using a curriculum learning~\cite{Bengio:ICML:2009} method to mitigate compounding errors arising from using model's own predicted answers. 

As described in \autoref{sec:intro}, we apply a variant of SS~\cite{bengio2015scheduled} to model the conversational history in CoQA during training by forcing the model to learn both from ground-truth answers and it's own predicted answers for the previous questions. 
The primary goal of this study is to develop a system that completely eliminates the reliance on ground-truth answers of previous questions and uses the model's own predicted answers for answering questions at test time.

This study is of crucial significance since none of the studies described above have investigated the problem of exposure bias in CoQA. 
Currently, there are no systems that solve exposure bias specifically in the context of CoQA. 
To the best knowledge of the authors, this is the first study which investigates the problem of exposure bias in the context of CoQA and proposes a SS-based method to overcome compounding errors in CoQA.

\section{Method} 
\label{sec_proposed_method}

\begin{figure*}[t]
    \centering
    \includegraphics[width=0.7\textwidth]{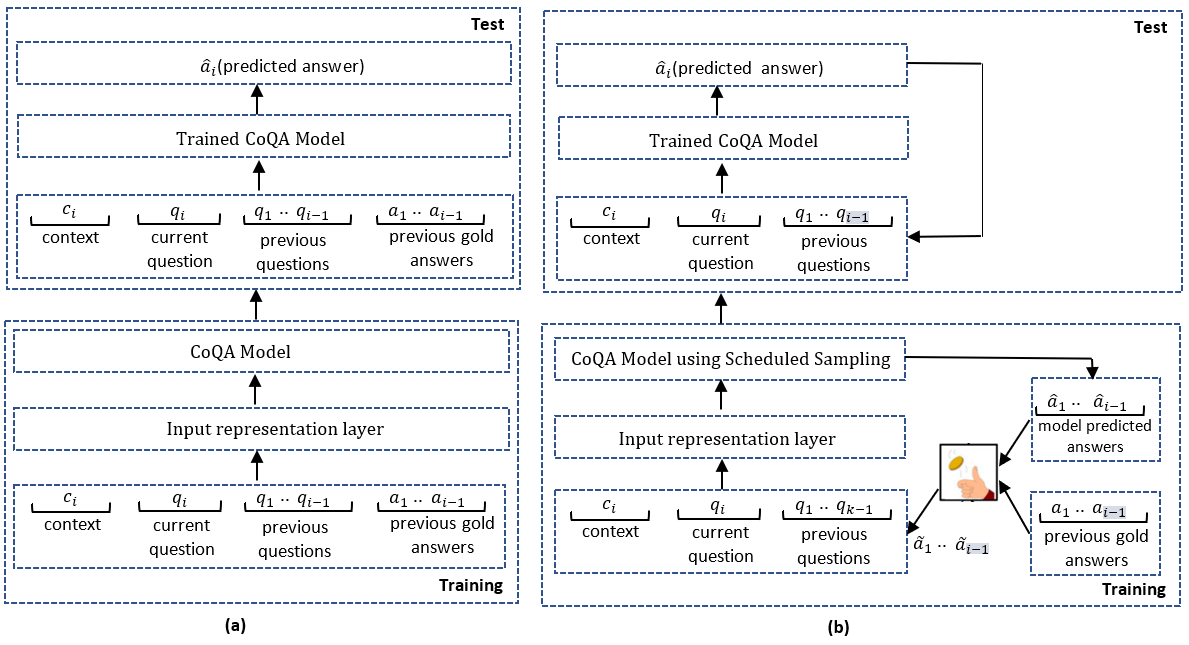}
    \caption{(a) Standard training method for CoQA. (b) Proposed schedule sampling method for CoQA.}
    \label{fig:method_diagram}
\end{figure*}

Formally, the task of CoQA is defined as follows. Given a passage $c_i$, and a set of previous questions $q_1, \ldots, q_{i-1}$ and answers $a_1, \ldots, a_{i-1}$, that provides conversational history, the task is to predict a response $\hat{a_i}$ for a given question $q_i$. 
A sampling mechanism is proposed for CoQA that randomly decides during training, whether to use ground-truth $a_1,.., a_{i-1}$ or model's predicted answers $\hat{a_1}, \ldots, \hat{a}_{i-1}$ for the previous questions. 
At test time, to predict the correct answer $\hat{a_i}$ for a test question $\hat{q_i}$, the model uses predicted answer utterances $\hat{a_1},.., \hat{a}_{i-1}$ instead of ground-truth answers $a_1, \ldots, a_{i-1}$ available in the test set. 
A schematic representation of the proposed method is shown in \autoref{fig:method_diagram}(b), in contrast to the standard training method used in CoQA, shown in \autoref{fig:method_diagram}(a).

As seen in \autoref{fig:method_diagram}(a), the contextual paragraph, $c_i$, the current question, $q_i$, previous questions, $q_1,.., q_{i-1}$, and the previous answer utterances, $a_1,.., a_{i-1}$, are used as inputs to train the CoQA model. 
Using a representation layer, the inputs are transformed into low dimensional vectors using pre-trained word embeddings such as GloVe \cite{pennington2014glove}, as used in Seq2Seq, PGNet, DrQA, DrQA+PGNet \cite{reddy2018coqa}, FlowQA \cite{huang2018flowqa} and SDNet \cite{zhu2018sdnet} models. 
Contextual embeddings such as ELMo \cite{peters2018deep} and CoVE \cite{mccann2017learned} are also used in FlowQA \cite{huang2018flowqa} and BERT \cite{devlin2018bert} is used in SDNet \cite{zhu2018sdnet} model. In our experiments, we use pre-trained GloVe \cite{pennington2014glove} and BERT \cite{devlin2018bert} embeddings with SDNet.


The word representation is then provided as the input to a reasoning layer, which primarily comprises of numerous sequence based networks such as LSTMs \cite{hochreiter1997long} or self-attention \cite{vaswani2017attention} to encode the context and question. 
The final output vector from the reasoning layer is then passed on to the answer prediction layer, which outputs the answer.
It includes (a) start and end indices of the most probable words in the contextual (answer span), and the probabilities of whether the answer is ``yes'', ``no'' or ``unknown'', for closed (yes/no) questions.
During inference, given $c_i$, $q_1, \ldots, q_{i-1}$ and $a_1, \ldots,  a_{i-1}$, the trained CoQA model is used to predict the answer $\hat{a}_i$ for the current question $q_i$ with probability  $p(\hat{a}_i|q_i)$ given by \eqref{eq:p-hat}.

\begin{align}
	\label{eq:p-hat}
   	p(\hat{a}_i|q_i) = f(c_i, q_{i}, a_{i-1}, q_{i-1}, \ldots, a_{1}, q_{1}) 
\end{align}
Therein, $f$ is a function used in existing CoQA models such as BiDAF++ \cite{yatskar2018qualitative}, FlowQA \cite{huang2018flowqa}, or SDNet \cite{zhu2018sdnet} for scoring the candidate answers. 
The candidate answer with the highest $ p(\hat{a}_i|q_i)$ value is predicted by the model, which can be either an answer span in the contextual paragraph or answers such as ``yes'' or ``no'' or ``unknown'', in the case of closed questions.

In contrast to the standard training procedure for CoQA (Figure \ref{fig:method_diagram}(a)) discussed above, the proposed SS method for CoQA (Figure \ref{fig:method_diagram}(b)) employs a  sampling technique that randomly selects during training whether we use ground-truth answers, $a_{i-1} \ldots, a_{1}$, or model's predicted answers, $\hat{a}_{i-1} \ldots, \hat{a}_{1}$, for the previous questions as the conversational history. 
For a given question $q_i$, where $i > 1$, we propose to flip a coin and choose for each $q_{i-1}$, the ground-truth answer $a_{i-1}$ with probability $\epsilon$, or the model's predicted answer $\hat{a}_{i-1}$ with probability $(1-\epsilon)$ as given in \eqref{eq:samp}.
\begin{align}
	\label{eq:samp} 
    \epsilon p(a_{i}|q_{i}) + (1-\epsilon) p(\hat{a}_{i}|q_{i}) ,
\end{align}
where $k$ is the total number of questions in a conversation. 
Intuitively, at the beginning of the training, since the model is not sufficiently trained, its predicted answers are likely to be incorrect and we must prefer the ground-truth answers. Later on, when the model becomes more accurate, we can shift towards selecting its predicted answers to simulate the scenario at test time, where we do not have access to ground-truth answers.
This can be achieved by adjusting the sampling rate, $\epsilon_{t}$, at epoch $t \in \{1, 2, \ldots, N\}$ following a schedule that depends on $t$, where $N$ is the maximum number of epochs.
We propose two such SS schemes for CoQA:
\begin{description}
  \item Uniform Sampling Rate (\textbf{USR}): This method applies a constant sampling rate $c \in \R$ for all epochs as given by \eqref{eq:USR}.
  \begin{align}
  	\label{eq:USR}
  	& \!\!\!\!\!\!\! \epsilon_t = c \quad \forall t \in  \{1, 2, \ldots, N\} &
  \end{align}
    \item{Exponential decay (\textbf{ED}):} This method exponentially varies the sampling rate as in \eqref{eq:ED}.
    \begin{align}
    	\label{eq:ED}
	\ \ \ \ \ \  \epsilon_t = 1-\exp(-t/2N) \quad \forall t \in  \{1, 2, \ldots, N\}    
    \end{align}    
  \end{description}

Let us denote the answer chosen for previous questions $q_{j}$ ($j\!<\!i$) by $\tilde{a}_{j}$.
The conversational history, $\tilde{a}_{i-1}, q_{i-1}, \ldots, \tilde{a}_{1}, q_{1}$, is provided as the input to the model. 
During inference, the model trained using SS uses $\hat{a}_{i-1}, q_{i-1}, \ldots, \hat{a}_{1}, q_{1}$  to predict answer $\hat{a}_i$ for the current question $q_i$ using the probabilities given in \eqref{eq:test-prob}.
This is in contrast to the standard training method that uses the ground-truth answers in the history, $a_{i-1}, q_{i-1}, \ldots, a_{1}, q_{i}$, for predicting answers
as given in \eqref{eq:test-prob}.
\begin{align}
	\label{eq:test-prob}
    p(\hat{a}_i|q_i) = f(c_i,q_i, \hat{a}_{i-1}, q_{i-1}, \ldots, \hat{a}_{1}, q_{1}) 
\end{align}

\section{Experiments}


\subsection{Dataset and Evaluation Metrics}

We conduct experiments using the CoQA dataset \cite{reddy2018coqa}. 
Since the test set of CoQA dataset is not publicly available, we report experimental results on development set. 
Further, as the main objective of this paper is to investigate the exposure bias in CoQA task and not to develop state-of-the-art for CoQA, the results obtained using the development set should be sufficient to draw conclusions with respect to the objectives of this study. 
Therefore, all the results reported in this paper pertains to the experiments carried out using the development set. 
We mainly report F1-scores to compare our results with previous work.

\subsection{CoQA Model}
\label{sec:coqa-model}

Although several systems are available for the CoQA task, we conduct experiments using SDNet \cite{zhu2018sdnet} because: 
(a) it achieves reasonable performance against state-of-the-art on CoQA; 
and (b) its implementation is made publicly available\footnote{\url{https://github.com/microsoft/SDNet}}. 
Although there are other systems that outperform SDNet, the non-availability of code restricts us in using them. 
However, it needs to be noted that although SDNet is used in our experiments as a specific instance of a CoQA method, 
the proposed SS methods are independent of SDNet and and can be applied to any CoQA method in principle.
Specifically, we experimented with SDNet\_Single \cite{zhu2018sdnet} model, denoted by \textbf{CoQAM}, under the following settings:
\begin{description}
    \item{\SM:} The CoQAM model is trained using the standard maximum likelihood, where ground-truth answers for previous questions are used both for training and at test time.
    \item{\MP:} Similar to \SM, this CoQAM model is trained using only the ground-truth answers. However, during test time the model's prediction (MP) for previous questions are used instead of the ground-truth answers as in \SM.
    \item{\SS[]:} The CoQAM model is trained using the proposed SS methods, as described in \autoref{sec_proposed_method}
\end{description}

\subsection{Implementation}


The SDNet framework was extended to include the scheduled sampling module, which was implemented in PyTorch. For efficient sampling during training all predictions from PyTorch implementation of $\textsc{lstm}$ is stored. This strategy avoids slower $\textsc{lstmcell}$ module which is not optimised by the CuDNN GPU-accelerated library\footnote{\scriptsize{\url{https://devblogs.nvidia.com/optimizing-recurrent-neural-networks-cudnn-5/}}}. 
Once stored, we sample at the next epoch $\epsilon$ as discussed in \autoref{sec_proposed_method}

\section{Results and Discussion} \label{sec:results}

The experimental results are presented in this section. We first present results of using ground-truth against model's prediction at test time, followed by the application of SS on the CoQA dataset.


\subsection{Ground-truth vs. Model's Prediction}

The performance of \SM and \MP on CoQA development set is shown in \autoref{table:overall_results}. 
\begin{table*}[t]
\begin{center}
\begin{tabular}{l c c c c c c c c c c c c}
\toprule
& \multicolumn{2}{c}{Child.} & \multicolumn{2}{c}{Liter.} & \multicolumn{2}{c}{Mid-High.} & \multicolumn{2}{c}{News.} & \multicolumn{2}{c}{Wiki.} & \multicolumn{2}{c}{Overall} \\
 & EM & F1 & EM & F1 & EM & F1 & EM & F1 & EM & F1 & EM & F1 \\
\midrule
\SM & 66.9 & 75.5 & 65.6 & 73.4 & 65.0 & 73.6 & 69.3 & 76.7 & 72.8 & 80.5 & 67.9 & 76.0 \\
\MP & 63.3 & 71.6 & 60.9 & 68.8 & 60.7 & 68.8 & 64.1 & 71.5 & 69.6 & 77.2 & 63.7 & 71.6  \\
\bottomrule
\end{tabular}
\end{center}
\caption{F1-Scores of SDNet model using ground-truth (\SM) and model's predicted (\MP) answers on the CoQA development set.}
\label{table:overall_results} 
\end{table*}
We find that there is a significant decrease in the performance in \MP that uses model's predicted answers for previous questions (achieves an overall F1-Score of 71.6) against a higher F1-score of 76.6, achieved by \SM model that uses the ground-truth answers during test time. 
This result clearly show that the human-written ground-truth answers are extremely essential for CoQA. 
Therefore, replacing ground-truth answers with model's predicted answers leads to significant exposure bias in CoQA. 
Further, the lower F1-Scores of \MP consistently across all domains strongly underline that compounding errors is a problem that is not specific to a particular domain but is a domain-independent general issue.

The answers predicted by \SM and \MP for a sample paragraph in the CoQA development set is shown in \autoref{table:model_output}.
Here, we do not show the whole paragraph due to space constraints, but show excerpts from the paragraph, where necessary. 
The example paragraph is chosen from the domain of ``children's stories'' with the paragraph id \texttt{3s3amizx3u5byyycmcbyzyr2n63cdu} in the CoQA development set. Both \SM and \MP use answers for the previous two questions in order to answer the current question. 

\begin{table*}[t]
\begin{center}
\begin{tabular}{c l l l l}
\toprule
 &  & \multicolumn{2}{c}{{Predicted answer}} &  \\
\multicolumn{1}{c}{Turn ID} & \multicolumn{1}{c}{Question} & \multicolumn{1}{c}{\SM} & \multicolumn{1}{c}{\MP} & \multicolumn{1}{c}{Ground-truth Answer} \\
\midrule
1. & What is the story about? & young girl and her dog & young girl and her dog & A girl and a dog \\
2. & What were they doing? & trip into the woods & trip into the woods & Set on on a trip \\
3. & Where? & into the woods & the woods & the woods \\
4. & How did the girl feel? & scared & scared	& scared \\
5. & How about the dog? & little & little & acting very interested \\
6. & How did he feel? & was acting very interested & was thinking of turning back & He was interested \\
7. & in what? & the bushes & in the bushes & what was in the bushes \\
8. & What was it? & bear & a small brown bear & a bear \\
9. & What did it do? & looked up at the girl & looked up at the girl & rested in the bushes \\ 
10. & Did it notice the two? & no & no & Not really \\
11. & How did the girl & scared & scared & surprised \\
 &  and the dog feel? &  &  &  \\
12. & How did the bear react? & friendly & surprised & friendly \\
\bottomrule
\end{tabular}
\end{center}
\caption{Answers predicted by \SM and \MP for a sample paragraph in the CoQA development set.}
\label{table:model_output}
\end{table*}

As seen from \autoref{table:model_output}, both \SM and \MP output similar answers for $Q_1$ to $Q_5$. 
Both models output an incorrect answer ``\emph{little}'' for $Q_5$, as there is little difference between ground-truth and model's own predictions for $Q_3$ and $Q_4$, used to answer $Q_5$. 
For $Q_6$,\SM predicts almost a correct answer ``\emph{was acting very interested}'',  whereas \MP predicts an incorrect answer ``was thinking of turning back''. 
The ground-truth answer for $Q_6$ is ``\emph{He was interested}''. 
\SM receives correct answers (ground-truth) for $Q_4$ and $Q_5$ as the input to the model, and predicts a correct answer for $Q_6$. 
However, the \MP predicts an incorrect answer, which is almost opposite in meaning (``\emph{was thinking of turning back}'') because it received an incorrect answer (model's prediction) as the input for $Q_5$. 
The answer to $Q_6$ is obtained from the following excerpt in the para: ``\emph{... The girl was a little scared and was thinking of turning back, but yet they went on. The girl's dog was acting very interested in what was in ..}''. 
As it can seen from this excerpt, due to the presence of the context ``\emph{scared}'' and absence of the context ``\emph{acting very interested}'', \MP predicts the answer ``\emph{was thinking of turning back}'', which is the feeling of the girl and not the dog. 
A similar problem can be seen when answering $Q_{12}$. 
Although \SM predicts the correct answer ``\emph{friendly}'' using the correct ground-truth answers (``\emph{Not really}'' and ``\emph{surprised}''), the \MP outputs an incorrect answer ``\emph{surprised}'', using model's predicted answers ``\emph{no}'' and ``\emph{scared}'', for the preceeding questions.

The above results clearly show that using correct answers for proceeding questions provides the vital context required for correctly answering questions in a conversational setting. 
However, using model's predicted answers for previous questions results in compounding errors due to the incorrect answers for the previous questions being used as the input to the model. 
Next, we evaluate how the proposed SS methods alleviate the problems arising from such compounding errors. 

\subsection{SS techniques for CoQA}

\begin{figure}[t]
    \centering
    \includegraphics[width=0.475\textwidth]{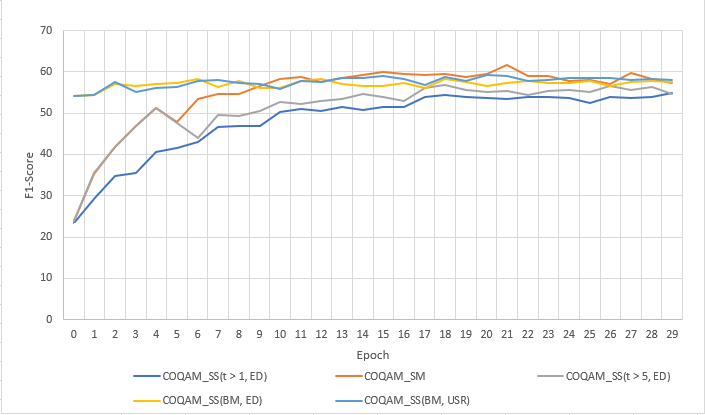}
    \caption{Performance of \SM against the different \SS[] methods that use different sampling techniques on ``Children's Stories'' domain in CoQA development set.}
    \label{fig:children_stories_performance}
\end{figure}

We explain below the results of applying SS for CoQA with an objective to overcome the exposure bias observed in CoQA. 
To this end, we evaluate different baseline methods that ED and USR at various stages of the training process as described next.
\begin{description}
    \item{\SS[$t > 1$, ED]:}. 
    This model applies \SS[] from the beginning of training using the exponential decay (ED) schedule described in \autoref{sec_proposed_method}
    Note that we must let the model to observe the ground-truth answer for each question at least once during training such that it can learn from the human-written answers. 
    Therefore, during the first epoch ($t = 1$), the model is trained using only the ground-truth answers as in \SM and the predictions for the training set are stored separately.      
     After the first epoch ($t > 1$), we sample between ground-truth and predicted answers for training the model further. 
     This model is trained for a total number of $N = 30$ epochs following~\newcite{zhu2018sdnet}.
     This model represents the performance of a CoQA model that had full supervision (teacher forcing) at a minimum level (only during the first epoch) and subsequently trained using a sampling scheme following ED.
     
      \item{\SS[BM, ED]:} 
     This model is trained using only the ground-truth answers, similar to the training phase of \SM, for 30 epochs to obtain the best model (BM) (measured on the validation data) that can be achieved by purely training using only the ground-truth answers.
    BM is likely to have a high exposure bias because it has not used at all model predicted answers during training.
     Next, the ED schedule is used to further train BM for another $N = 30$ epochs, allowing BM to recover from the exposure bias.
     This baseline will demonstrate the effectiveness of ED to fine-tune a CoQA model that is pre-trained purely using the ground-truth answers.
    
    \item{\SS[$t>5$, ED]:} This model is similar to \SS[$t > 1$, ED] except that ED begins after the 5-th epoch. In other words, the model is trained using only the ground-truth answers for the first 5 epochs, enabling it to fit to ground-truth answers for a longer duration than in \SS[$t>1$, ED].
    This model can be seen as a compromise between pre-training a CoQA model using only the ground-truth (thereby overfitting to ground-truth answers and inducing exposure bias) as in \SS[BM, ED] vs. not sufficiently exposing the model to the ground-truth answers (thereby underfitting the model) as in \SS[$t>1$, ED].
     
    \item{\SS[BM, USR]:}
    This method is identical to \SS[BM, ED] above, except that we use USR with a constant sampling rate of $c = 0.5$ during training instead of ED.
    This baseline will demonstrate the effectiveness of USR to fine-tune a CoQA model that is pre-trained purely using the ground-truth answers.
    
\end{description}
The experiments were conducted on the ``Children Stories'' domain in the CoQA dataset.
In all of the above experiments, during test time we use the predicted answer ($\hat{a}_i$) in the conversational history, and not the ground-truth answers ($a_i$). 
The performances of the above models on the CoQA development set are shown in \autoref{fig:children_stories_performance} and the F1-Scores achieved by these models are shown in \autoref{table:children_stories_f1_scores}.

\begin{table}[t]
\begin{center}
\begin{tabular}{l c}
\toprule
Model & F1-Score \\
\midrule
\multicolumn{2}{c}{Different Sampling Techniques} \\
\midrule
\SS[$t>0$, ED]  	& 54.8 \\
\SS[$t>5$, ED]  	& 54.7 \\
\SS[BM, ED]		& 58.2 \\
\SS[BM, USR]	& 59.1 \\

\smallskip \\
\midrule
\multicolumn{2}{c}{Ground-truth vs. Model's Prediction} \\
\midrule
\MP & 54.1 \\
\SM & 61.6 \\
\bottomrule 
\end{tabular}
\end{center}
\caption{F1-Scores achieved by different sampling techniques for the ``Children's Stories'' domain in the CoQA development dataset.}
\label{table:children_stories_f1_scores}
\end{table}

\subsection{Overcoming exposure bias in CoQA}

\autoref{table:children_stories_f1_scores} shows that \SM, which uses the ground-truth answers at test time, achieves the best F1-Score of 61.6. 
However, ground-truth answers will \emph{not} be available in real-world CoQA settings.
Therefore, \SM must be considered as an upper-bound performance that we could hope to obtain if we had access to ground-truth answers even during test time.
The \MP model which is trained exactly as \SM by using only the ground-truth answers but uses its own predictions during test time as the history of the conversation achieves a lower F1-Score of 54.1. 
This significant drop in performance from 61.6 to 54.1 is an indication of the exposure bias in the \SM model.

From \autoref{table:children_stories_f1_scores}, we see that the different SS  methods proposed to overcome the exposure bias substantially helps in mitigating the problem of compounding errors to different degrees.
Among the different sampling methods \SS[BM, USR] achieves the best F1-Score of 59.1. 
This is followed by \SS[BM, ED], which achieves an F1-Score of 58.2. 
As seen from \autoref{fig:children_stories_performance}, the methods that retrain the best model obtained first by training using only the ground-truth answers and then subsequently fine-tune using ED or USR achieve performance similar to that of \SM. 
However, \SS[$t>0$, ED] and \SS[$t>5$, ED], which respectively apply ED right from the beginning and at mid-way through, struggle to achieve a comparable performance against \SM. 
Recall that we first train all SS methods for one epoch using only the ground-truth answers. 
Therefore, even if we use a fixed sampling rate as in USR, by selecting a sufficiently small sampling rate, we can still expose the model to both ground-truth as well as the model's own predictions at training time to overcome exposure bias.
These results indicate that applying SS to fine-tune the best-CoQAM models is more useful in the context of CoQA.

In particular, we see that \SS[BM, USR], which fine-tunes the best CoQAM model using a uniform sampling rate achieves the highest F1-Score of 59.1. 
Even though  \SS[BM, USR] does not outperform \SM, the F1-score of 59.1 achieved by  \SS[BM, USR] is a significant improvement over the F1-Score of 54.8, achieved by \MP. 
 \SS[BM, USR] has a greater advantage in that it does not use any ground-truth answers during test time, which is a more realistic setting.   
The performance of this best performing  \SS[BM, USR] model on other domains in the CoQA dataset is shown in \autoref{table:performance_other_domains}. 

We see that there is a significant drop across different domains in the CoQA dataset when the model's predicted answer is used during inference due to exposure bias. 
Overall, applying \SS[BM, USR]) helps to consistently improve the F1-Scores across all domains. 
These results clearly show the usefulness of applying SS techniques in the context of CoQA for developing more robust CoQA models, which do not depend on the ground-truth answers and is able to use their own predictions to model the conversational history.

\begin{table}[t]
\centering
\resizebox{\columnwidth}{!}{%
\begin{tabular}{l c c c c}
\toprule
& Liter. & Mid-High. & News & Wiki  \\
\midrule
\SS[BM, USR] & 65.12 & 65.50 & 67.17 & 75.19\\
\MP & 63.98 & 64.20 & 65.11 & 74.42 \\
\SM & 69.69 & 67.63 & 71.14 & 77.95 \\
\bottomrule
\end{tabular}%
}
\caption{ F1-Scores of \SS[BM, USR] on various domains in the CoQA development set.}
\label{table:performance_other_domains}
\end{table}

\subsection{Performance on Question Types}


In CoQA task, the predicted answer is either a continuous span of text in the paragraph associated with a conversation, or a closed-form answer such as ``yes'', ``no'' or ``unknown''. 
\autoref{table:performance_question_types} shows the performance of the different models for various types of questions across different domains in the CoQA development set. 
From \autoref{table:performance_question_types} we see a significant decrease in performance when the model's own predictions are used during inference time (\SM vs \MP), particularly for questions that require continuous text spans as answers. 
This holds true across all domains in the CoQA dataset.
Moreover, we see that the proposed SS methods for CoQA mitigate the compounding errors in predicting answer spans across all domains. 
Interestingly, \SS[BM, USR] outperforms \SM~for  ``yes'' or ``no'' type questions. 
For example, for ``no''-type questions, \SS[BM, USR] performs well for the domains ``Literature'', ``Mid-High'' and ``Wikipedia'', whereas for ``yes''-type questions,   \SS[BM, USR] performs well for the domains ``Children's stories'', ``Mid-High'', ``News'' and ``Wikipedia''. 
These results indicate that SS methods are particularly effective when answering ``yes'' or ``no'' type questions.

\begin{table}[t]
\begin{center}
{\small
\begin{tabular}{p{0.6cm}  c  c c}
\toprule
 Type & \SS[BM, USR] & \MP &  \SM   \\ 
\midrule \midrule
\smallskip
 & \multicolumn{3}{c}{Children Stories} \\
\midrule
yes & 57.81 & 47.07 & 50.19 \\
no & 73.12 & 75.46 & 74.84 \\
unknown & 38.09 & 35.18 & 35.18 \\
span & 57.40 & 52.03 & 59.91 \\
overall & 59.12 & 54.14 & 60.61 \\
\midrule
 & \multicolumn{3}{c}{Literature} \\
\midrule
yes & 71.03 & 80.81 & 81.80 \\
no &  82.37 & 75.15 & 73.23 \\
unknown & 46.79 & 46.17 & 52.42 \\
span &  62.28 & 60.63 & 67.54 \\
overall & 65.12 & 63.98 & 69.69 \\
\midrule
& \multicolumn{3}{c}{Middle-High School Exams (Mid-High)} \\
\midrule
yes & 67.50 & 64.07 & 65.41 \\
no & 85.13 & 81.94 & 79.69 \\
unknown & 25.00 & 32.80 & 30.95 \\
span &  62.73 & 61.90 & 66.49 \\
overall & 65.50 & 64.20 & 67.63 \\
\midrule
& \multicolumn{3}{c}{News} \\
\midrule
yes & 58.15 & 47.63 & 48.15 \\
no & 83.84 & 84.03 & 86.17 \\
unknown & 33.33 & 33.33 & 38.33 \\
span & 66.36 & 64.62 & 71.45 \\
overall &  67.17 & 65.11 & 71.14 \\
\midrule
& \multicolumn{3}{c}{Wikipedia} \\
\midrule
yes &  72.19 & 72.19 & 71.06 \\
no & 91.83 & 83.01 & 83.45 \\
unknown & 49.36 & 49.81 & 56.48 \\
span & 74.02 & 73.99 & 78.08 \\
overall & 75.19 & 74.42 & 77.95 \\
\bottomrule
\end{tabular}
}
\end{center}
\caption{ F1-Scores reported by the best performing \SS[BM, USR] on different question types across domains in the CoQA development set.}
\label{table:performance_question_types}
\end{table}

\subsubsection{Effect of Conversation Length}

The problem of compounding errors is likely to be severe for longer conversations (with higher number of questions) as questions towards the end of conversations can potentially contain many incorrectly predicted answers during test time. To study the effect of conversation length on the performance of a CoQA model we group the conversations in the CoQA dataset into two categories based on the number of questions in the conversation:

\begin{itemize}
    \item [a.] \textbf{shorter conversations} that contain less than 12 rounds of questions and answers (a question and its answer forms a single round)
    \item [b.]\textbf{longer conversations} that contain more than 12 rounds
\end{itemize}

The experiments were conducted on conversations in the ``Children's Stories'' domain of CoQA development set.

  As seen from \autoref{table:performance_conv_history}, \SS[BM, USR] performs significantly better than \SM for shorter conversations -- particularly for question types ``yes'' (73.33 vs. 70.55); ``no'' (52.70 vs. 47.29); and ``unknown'' (42.85 vs. 37.63). 
  For longer-conversations, \SS performs better (59.89) than \SS[BM, USR] (51.37), especially for ``no'' type questions. 
  The lower performance of \MP compared to  \SM in all question types (except ``yes'' type) for both shorter and longer conversations indicates the presence of exposure bias across all question types. 
  The above results shows that  \SS[BM, USR]  is more useful for shorter conversations.
  Proposing methods to overcome compounding errors for longer conversations remains a challenge for future work. 

\begin{table}[t]
\begin{center}
{\small
\begin{tabular}{p{0.6cm} c c c}
\toprule
&  \SS[BM, USR] &  \MP & \SM  \\
\midrule
\midrule
 & \multicolumn{3}{c}{Shorter Conversations} \\
\midrule
yes & 73.33 & 83.88 & 70.55 \\
no & 52.70 & 44.59 & 47.29 \\
unknown & 42.85 & 37.63 & 37.63 \\
span &  55.85 & 48.69 & 57.08 \\
overall & 57.23 & 51.74 & 57.48 \\
\midrule
 & \multicolumn{3}{c}{Longer Conversations} \\
\midrule
yes & 73.04 & 72.17 & 76.52 \\
no & 59.89 & 48.07 & 51.37 \\
unknown & 33.33 & 33.33 & 33.33 \\
span & 58.16 & 53.67 & 61.29 \\
overall & 60.02 & 55.28 & 62.08 \\
\bottomrule
\end{tabular}
}
\end{center}
\caption{\label{table:performance_conv_history} F1-Scores of models with conversation length.}
\end{table}

\subsection{Effect of Question Length}


Longer questions are likely to be complex and difficult to correctly answer. 
To empirically evaluate the relationship between the length of a question, $QL$, (measured by the number of tokens in the question) and the accuracy of a CoQA model, we conduct the following experiment.
First, we categorise the questions in the ``Children's stories'' domain into five categories: (a) $QL_{1}:  QL < 3$; (b) $QL_2: 3 \leq QL < 5$; (c) $QL_3: 5 \leq QL < 7$; (d) $QL_4: 7 \leq QL < 10$; (e) $QL_5: \geq 10$.


The performance of the different CoQA models on the question types are shown in \autoref{table:performance_question_length}.
A dash indicates that there were no questions satisfying the length requirement in that category or type.
From \autoref{table:performance_question_length} we see that \SS[BM, USR] performs better than \SM for questions in category $QL_3$ (overall score of 62.04 vs. 60.97) and $QL_5$ (overall score of 62.22 vs. 60.24), indicating that SS method can be more useful for longer questions. 
On the other hand, for shorter questions $(QL_1 \& QL_2)$, \SM performs better than \SS[BM, USR] as the former has access to ground-truth answers in spite of shorter questions.

\begin{table}[t]
\begin{center}
\begin{tabular}{l  c  c  c  c  c}
\toprule
 & $QL_1$ & $QL_2$ & $QL_3$ & $QL_4$ & $QL_5$\\ 
 \medskip \\
& \multicolumn{5}{c}{\SS[BM, USR]} \\
\midrule
yes & - & 66.66 & 79.10 & 70.49 & 50.00 \\
no & - & 66.66 & 49.07 & 63.00 & 100.00 \\
unknown & - & - & - & 57.14 & -  \\
span & 48.99 & 48.23 & 61.26 & 57.62 & 59.78 \\
overall & 48.18 & 52.34 & 62.04 & 59.67 & 62.22 \\
\smallskip \\ \midrule
& \multicolumn{5}{c}{\MP} \\
\midrule
yes & - & 73.33 & 71.26 & 80.32 & 100.00 \\
no & - & 42.85 & 41.66 & 53.50 & 100.00 \\
unknown & - & - & - & 52.77 & -  \\
span & 38.82 & 47.38 & 54.24 & 53.19 & 52.85  \\
overall & 38.19 & 50.31 & 54.91 & 56.45 & 60.40 \\
\smallskip \\ \midrule
& \multicolumn{5}{c}{\SM} \\
\midrule
yes & - & 76.66  & 74.25 & 75.40 & 50.00 \\
no & - & 42.85 & 39.81 & 63.50 & 100.00 \\
unknown & - & - & - & 52.77 & -  \\
span & 64.41 & 59.34 & 61.61 & 57.49 & 57.43 \\
overall &  63.35 & 59.87 & 60.97 & 60.18 & 60.24 \\
\bottomrule
\end{tabular}
\end{center}
\caption{ F1-Scores of models on different question lengths.} 
\label{table:performance_question_length}
\end{table}

\section{Conclusion}
In this paper we examined the problem of exposure bias in the context of CoQA and proposed the use of a sampling-based method to effectively mitigate it in the context of CoQA. The empirical results and analysis provided clearly show the presence of exposure bias in CoQA. Experiments with different sampling rates shows that SS is indeed helpful in developing CoQA models, which are more useful in real-world scenarios, where human-written ground-truth answers are not available for predicting answers. From this we conclude that future evaluations of CoQA systems should not be given predicted answers instead of ground-truth answers at test time.

\section{References}

\bibliographystyle{lrec}
\bibliography{myrefs}


\end{document}